\documentclass[11pt]{article}

\usepackage{macros}
\usepackage{amsmath} 
\usepackage{graphicx}
\usepackage{units}
\usepackage{pstricks}
\usepackage{amssymb}
\usepackage{times}

\begin{document}

\begin{center}
  \begin{flushleft}
    DESY 06-214 \\
    November 2006
  \end{flushleft}
  \vspace{1cm}
  \begin{Large}
    {\bf Bottom production cross section from double muonic decays of } \\
    {\bf \boldmath{\bfl}~hadrons in \boldmath 920 \bf GeV proton-nucleus collisions } \\
  \end{Large}
  \vspace{0.5cm}
  \begin{small}
    I.~Abt$^{23}$,
M.~Adams$^{10}$,
M.~Agari$^{13}$,
H.~Albrecht$^{12}$,
A.~Aleksandrov$^{29}$,
V.~Amaral$^{8}$,
A.~Amorim$^{8}$,
S.~J.~Aplin$^{12}$,
V.~Aushev$^{16}$,
Y.~Bagaturia$^{12,36}$
% ...

\vspace{5mm}

% ...
$^{8}${\it LIP Coimbra, P-3004-516 Coimbra,  Portugal~$^{f}$} \\
% ...
$^{10}${\it Institut f\"ur Physik, Universit\"at Dortmund, D-44221 Dortmund, Germany~$^{d}$} \\
% ...
$^{12}${\it DESY, D-22603 Hamburg, Germany} \\
$^{13}${\it Max-Planck-Institut f\"ur Kernphysik, D-69117 Heidelberg, Germany~$^{d}$} \\
% ...
$^{16}${\it Institute for Nuclear Research, Ukrainian Academy of Science, 03680 Kiev, Ukraine~$^{h}$} \\
% ...
$^{23}${\it Max-Planck-Institut f\"ur Physik, Werner-Heisenberg-Institut, D-80805 M\"unchen, Germany~$^{d}$} \\
% ...
$^{29}${\it DESY, D-15738 Zeuthen, Germany} \\
% ...
$^{36}${\it visitor from High Energy Physics Institute, 380086 Tbilisi, Georgia} \\
% ...

  \end{small}
\end{center}

\begin{abstract}
  The \bbbar{} production cross section in $920$ GeV proton-nucleus 
  fixed target collisions is measured by observing double muonic 
  decays of \bfl{} hadrons in the kinematic region 
  $-0.3<\xF(\mu)<0.15$.
  A total number of $76\pm12$ \bbbar{} events is obtained with a 
  likelihood fit of the signal and background simulated events to 
  the data.
  The resulting cross section is 
  $\sigbb = 16.2 \pm 2.5_{stat} \pm 2.8_{sys}~\nbpn$, or, when 
  combined with a previous \HeraB{} measurement of similar precision, 
  $\sigbb = 15.4 \pm 1.7_{stat} \pm 1.2_{sys}^{uncorr.} \pm 1.9_{sys}^{corr.}~\nbpn$,
  which is consistent with recent NLO calculations.
\end{abstract}

  \twocolumn
\section{Introduction}

The measurement of bottom production in fixed target collisions 
offers the possibility to test perturbative QCD in the 
near threshold energy regime, where the effect of higher order 
processes, such as soft gluon emission, has been calculated 
\cite{Bonciani:1998vc, Kidonakis:2004qe}.
At first order, the production mechanism at the \HeraB{} energy 
(\cms) is dominated by gluon-gluon fusion (\ggtobb) 
\cite{Kidonakis:2001nj}.

Three experimental results are published.
Two are inconsistent, even though they were obtained by similar 
experiments searching for \Jpsi{} \cite{Jansen:1994bz} and 
semi-leptonic \cite{Alexopoulos:1997zx} decays of \bfl{} hadrons.
\HeraB{} recently published the most accurate result based on a 
measurement of \Jpsi{} decays of the \bfl{} hadrons \cite{Abt:2005qs}.

In this paper, a measurement of the \bbbar{} production cross section 
performed with a $b$ tagging technique independent from our previous 
measurement is presented.

After production, \bbbar{} pairs hadronise and mostly decay into 
\cfl{} hadrons.
Since \bfl{} and \cfl{} hadrons have a large probability to decay 
with the emission of a muon (``semi-muonic decay'') \cite{PDG}, the 
\bbbar{} production cross section is measured by searching for 
\bbtommX{} decay events, in which at least two of the four heavy 
quarks typically produced in a \bbbar{} event ($b$, \bbar, $c$, 
\cbar) undergo semi-muonic decays (``double muonic $b$ decays'').

The \bbbar{} event selection is based on a pair of oppositely charged 
muons not coming from the primary interaction vertex, having a large 
momentum transverse to the beam.

\section{Detector and data sample}

\HeraB{} \cite{Lohse:9402, Hartouni:9501} is a large acceptance 
forward spectrometer installed at the $920$ GeV proton storage ring 
of DESY.
The Feynman-$x$ (\xF) of accepted muons from semi-muonic $b$ decays 
is between $-0.3$ and $0.15$.
The detector is used to reconstruct charged particle tracks produced 
in the interactions of the proton beam halo with wires of different 
materials (\Carbon, \Titanium{} and \Tungsten), in several 
configurations \cite{Ehret:2000iy}.
Particles are tracked with a silicon microstrip detector 
\cite{Bauer:2003rt} whose first station (of $8$) is a few centimeters 
from the target system and which extends approximately $2~$m further 
downstream. 
A primary vertex resolution of $500~\mu$m along the beam and 
$50~\mu$m in the perpendicular plane is achieved.
Up to $13~$m downstream of the target, honeycomb chambers in the 
outer region \cite{Albrecht:2005qt, Albrecht:2005qu}, and microstrip 
gaseous chambers in the inner region \cite{Bagaturia:2002}, allow to 
track particles and to measure their momenta from the bending in a 
$2.13$ T$\cdot$m vertical magnetic field. 
A Cherenkov detector \cite{Arino:2003in} is used for $\pi/K/p$ 
separation.
An electromagnetic shashlik calorimeter \cite{Avoni:2001si} serves 
for $e$ and $\gamma$ identification.
At the rear of the detector, muons with momenta larger than $5$ 
\GeVpc{} are tracked with triple stereo layers of gaseous tube 
chambers interleaved with hadron absorbers \cite{Arefiev:2002}.

\begin{figure}[htb]
  \centerline{\includegraphics[width=0.5\textwidth]
    {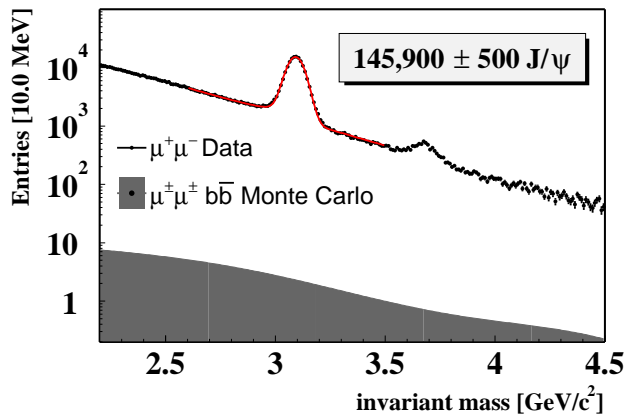}}
  \caption{Invariant mass of opposite-sign muons after dimuon 
    selection (white histogram).
    Two peaks emerge above the background (\Jpsi{} and \Psip).
    The fit of the \Jpsi{} signal at $3.097$ \GeVpcc{} (grey line) 
    includes events in the radiative tail \cite{Spiridonov}.
    The grey histogram is a Monte Carlo simulation of double muonic 
    $b$ decays (see text for details).}
  \label{fig:jpsi}
\end{figure}

Double muonic $b$ decays are searched for in 
$164\negthinspace\cdot\negthinspace10^6$ events of $p$-C, $p$-Ti and 
$p$-W interactions ($1.4$ overlapped interactions per event, in average) 
collected with a multilevel dilepton trigger \cite{Balagura:2002kr} in 
the $2002$-$2003$ data taking period.
The trigger is designed to select dilepton decays of \Jpsi{} mesons 
produced in the proton-nucleus collisions.
By applying dimuon selection criteria similar to those of reference 
\cite{Abt:2005qs}, about $146,\negthinspace000$ prompt \Jpsi{} mesons 
are reconstructed.
With this event selection, double muonic $b$ decays are a tiny 
fraction of the surviving muon pairs (Figure \ref{fig:jpsi}).

The systematic uncertainties due to the detector performance and 
acceptance is reduced by normalising the \bbbar{} production cross 
section to the prompt \Jpsi{} cross section.

\section{Monte Carlo simulation}

The acceptance and the reconstruction efficiencies are determined 
through Monte Carlo simulations of the physics processes occurring in 
proton-nucleus interactions.
For bottom and charm production, PYTHIA 5.7 \cite{Sjostrand:1993yb} 
is used.
For bottom production, the predictions of NRQCD models are used, 
while \Jpsi{} production is tuned to match the \xF{} and 
\pT{} distributions measured by other experiments.
The fragmentation process is simulated by JETSET 7.4 
\cite{Sjostrand:1993yb}.
The energy left from the hard scattering is used by FRITIOF 
\cite{Pi:1992ug} to simulate the underlying inelastic event.
The response of the detector is simulated by GEANT 3.21 
\cite{Brun:1987}.
See, for more details, reference \cite{Abt:2005qs}.

\section{Measurement method}

The \bbbar{} production cross section can be expressed as 

\begin{displaymath}
  \sigbb = \frac{\Nbbpm \Dsigjp \BR{\Jpsi}}{\Xbbpm}, 
\end{displaymath}
where \Nbbpm{} is the total number of double muonic $b$ decays in our 
\mpmm{} sample, $\Dsigjp = 417 \pm 37~\nbpn$ \cite{Maltoni:2006yp} is 
the prompt \Jpsi{} production cross section in the detector 
acceptance (\mbox{$-0.35<\xF(\Jpsi)<0.15$}) and \BR{\Jpsi} is the 
branching ratio for \jptomm{} decays ($5.88 \pm 0.10\%$) \cite{PDG}.
The term \Xbbpm{} is defined as 

\begin{displaymath}
  \Xbbpm = \sum_{i} \Njpi A^{1-\alpha}_{i} \frac{\sum_{j} \BR{Bj} (1-\thj) \ebbpmji}{\ejpi},
\end{displaymath}
where the index $i$ runs over $14$ different target configurations.
The index $j$ refers to the different ways to generate dimuons via 
semileptonic $b$ or \btoc{} decays (Section \ref{sec:signal}) having 
different branching ratios (\BR{Bj}) and reconstruction efficiency 
(\ebbpmji, the superscript indicates the dimuon charge).
The charge factors \thj{} gives the fraction of resulting same-sign 
muons (Table \ref{tab:branching}), \Njpi{} is the number of \Jpsi{} 
mesons reconstructed with efficiency \ejpi, $A$ is the atomic weight 
of the target and \mbox{$\alpha = 0.96\pm0.01$ \cite{Leitch:1999ea}} 
is the \Jpsi{} nuclear suppression in the central \xF{} region.
Nuclear effects are expected to be negligible for open bottom and 
open charm production.

The number \Nbbpm{} is obtained with two methods leading to two 
determinations of the \bbbar{} production cross section (Section 
\ref{sec:analysis}).

\section{Signal decay modes \label{sec:signal}}

The signal sample consists of events with two muons coming from 
semi-muonic decays of heavy quarks.
Four heavy quarks are typically produced in a \bbbar{} event ($b$, 
\bbar, $c$, \cbar).
Depending on the type of hadrons decaying in the semi-muonic mode 
(\bibj, \bici, \bicj{} or \cicj), four classes ($Bj$) of signal 
events are defined \cite{Sbrizzi:2006jx} (Table \ref{tab:branching}).

\vspace{0.3cm}

\begin{table}[htb]
  \center
  \begin{tabular}{|c|c|c|c|}
    \hline
    Class & Decaying hadrons & Branching ratio & $\theta$ \\
    \hline
    $B1$ & \bibj  & $0.0084$ & $0.187$ \\
    $B2$ & \bici  & $0.0160$ &     $0$ \\
    $B3$ & \bicj  & $0.0156$ & $0.742$ \\
    $B4$ & \cicj  & $0.0081$ & $0.315$ \\
    \hline
  \end{tabular}
  \caption{Branching ratios and charge factors $\theta$ (see text 
    for details) of four classes of double muonic $b$ decays. 
    The branching ratios are correlated and are affected by a 
    relative uncertainty of $10\%$.
    The relative uncertainty on $\theta$ is less than $1\%$.
  }
  \label{tab:branching}
\end{table}

For each class, the branching ratio for dimuon decays is obtained 
from the branching ratios for semi-muonic $b$ and $c$ decays reported 
in the PDG \cite{PDG}.
The fraction of decays into same-sign muons (\mpmp) is given by a 
charge factor $\theta$, \mbox{$1$-$\theta$} being the fraction of 
decays into opposite-sign muons (\mpmm).
The value of $\theta$ is determined through Monte Carlo simulations 
and includes the effect of \Bzero{} and $B_S$ mixing and all possible 
decay paths leading to muon pairs.

A fifth class, which is not included above, must be also considered.
Events in which \mpmm{} pairs originate from double muonic decay of 
\ccbar{} pairs from \btoccsX{} decays represent $5.5\%$ of the total 
signal events.
These events are assigned to the class $B2$, which exhibits a similar 
final state.
The systematic uncertainty corresponding to this choice is included 
in the branching ratio \Br{\btoccsX}.

Events with more than two muons might fall into more than one class.
In order to avoid multiple counting of signal events, the calculated 
branching ratios in Table \ref{tab:branching} include the probability 
that \bfl{} and \cfl{} hadrons do not decay into muons, and events 
are assigned to classes with a priority given by the order of classes 
in Table \ref{tab:branching}.

\section{Background contributions \label{sec:background}}

The main background for \bbbar{} event selection is made of events 
having two muon-like tracks not coming from the primary interaction 
vertex.
Two sources of this type are considered: double muonic decays of 
\cfl{} hadrons and random combinations of muons from decay of low 
mass mesons (mainly pions and kaons).
The latter background is referred to as ``combinatorial background''.

The Monte Carlo simulation shows that a cut around the \Jpsi{} mass 
(between $2.95$ \GeVpcc{} and $3.25$ \GeVpcc) removes background 
events from \btojpX{} decays and prompt \Jpsi{} decays.
The \Jpsi{} mass region is excluded to be also statistically 
independent from the measurement of \bbbar{} production cross section 
in \cite{Abt:2005qs}, where \bbbar{} events are identified with 
\Jpsi{} decays of the \bfl{} hadrons.

Muons from Drell-Yan events are at least $10$ times less abundant 
than the combinatorial background in the invariant mass region of 
interest \cite{Abt:2006wc}.

The number of background events from double muonic $c$ decays 
(\Nccpm) is determined through a Monte Carlo simulation normalised to 
the number of prompt \Jpsi{} mesons reconstructed in the data.
It can be expressed as 

\begin{displaymath}
  \Nccpm = 
  \frac{\sigcc \BR{\ccbar}}{\Dsigjp \BR{\Jpsi}} \Xcc,
\end{displaymath}
where \sigcc {} is the charm cross section \cite{Lourenco:2006vw} 
($49 \pm 5~\mubpn$ \cite{charm}), 
\mbox{$\BR{\ccbar}=\Br{\ctomX}^2=(0.082\pm0.005)^2$} 
\cite{PDG}, and the term \Xcc{} is defined as 

\begin{displaymath}
  \Xcc = \sum_{i} \Njpi A^{1-\alpha}_{i} \frac{\ecci}{\ejpi}, 
\end{displaymath}

The term \ecci{} is the \bbbar{} selection efficiency for 
double muonic $c$ decay events in the target configuration $i$.

The combinatorial background in the \mpmm{} channel (\Ncopm) is 
determined with \mpmp{} data.

Muons from double muonic $b$ decays do not come from the same decay 
vertex.
However, the forward boost in fixed target collisions is such that 
tracks coming from two long lived particle decays are almost as close 
as those originating from a single particle decay.
In double muonic $b$ decays, the middle point of the segment of 
minimum distance between the two muons ($pmd$) is preferentially 
located downstream of the target, while the region upstream of the 
target is dominated by combinatorial background, in both \mpmm{} and 
\mpmp{} channels.

Assuming that the combinatorial background in the two final states 
has a similar shape, the number of events are normalised with respect 
to the upstream side and the combinatorial background in the 
\mpmm{} channel is estimated as the difference between the number of 
\mpmp{} pairs in data (\Nmmpp) and those expected from double muonic 
$b$ decays in the same channel (\Nbbpp) 

\begin{displaymath}
  \Ncopm = \Nmmpp - \Nbbpp.
\end{displaymath}

\Nbbpp{} is obtained from Monte Carlo simulations, under the 
assumption that \mbox{$\sigbb = 15~\nbpn$}, with the formula 

\begin{displaymath}
  \Nbbpp = \frac{\sigbb}{\Dsigjp \BR{\Jpsi}} \Xbbpp,
\end{displaymath}
where the term \Xbbpp{} is defined as 

\begin{displaymath}     
  \Xbbpp = 
  \sum_{i} \Njpi A^{1-\alpha}_{i} \frac {\sum_{j} \BR{Bj} \thj \ebbppji}{\ejpi}.
\end{displaymath}

The term \ebbppji{} indicates the reconstruction efficiency for 
double muonic $b$ decays into \mpmp.

\section{Data analysis \label{sec:analysis}}

Two methods have been used to estimate the number of events from \bbbar{} 
decay.  
A first method, which is used to obtain our final results, is described 
in Section \ref{sec:method1} and a second method, which is used as a 
cross-check, is described in Section \ref{sec:method2}.
With one exception (discussed below), the choice of cuts used in both 
methods is the same.
Initial muon selection cuts are given in Section \ref{sec:muon} and 
the procedure for optimizing the final cuts is described in Section 
\ref{sec:cutopt}.

The selection of \bbbar{} decay events begins by requiring that the 
events have at least two muons.

\subsection{Muon selection \label{sec:muon}}

A first general muon selection is performed by requiring a 
high-quality reconstructed triggered track having a momentum between 
$5$ and $200$ \GeVpc, a minimum transverse momentum (\pT) of 
$0.7$ \GeVpc{} and a minimum \chisq{} probability of the track fit 
($P_{\chi}$) of $0.003$.
The muon likelihood \cite{Titov:2000eb}, as measured in the muon 
detector, must be greater than $0.05$.

\subsection{Cut optimisation \label{sec:cutopt}}

Both muons of an event are required to have a minimum \pT{} of 
$1~\GeVpc$ and a minimum impact parameter to the target ($Ip$) of 
$1.5\sigma$ (where $\sigma$ is the $Ip$ resolution).
The $Ip$ is defined as the perpendicular distance between the target 
wire and the point on the track extrapolated to the $z$-position of 
the target. 
The dimuon invariant mass ($m$) is required to be at least 
$2$ \GeVpcc.

The optimal \bbbar{} selection criteria are found by maximising the 
signal significance $S$, which is defined as 

\begin{displaymath}
  S = \frac{\Nbbpm}{\sqrt{\Nbbpm + \Nccpm + \Ncopm}}.
\end{displaymath}

The number of signal events (\Nbbpm), which consists of oppositely 
charged muons, is obtained from Monte Carlo simulations, under the 
assumption that \mbox{$\sigbb = 15~\nbpn$}:

\begin{displaymath}
  \Nbbpm = \frac{\sigbb}{\Dsigjp \BR{\Jpsi}} \Xbbpm.
\end{displaymath}

The formulae used to estimate the number of background events 
(\Nccpm{} and \Ncopm) have been presented in Section \ref{sec:background}.

A large fraction of combinatorial background consists of muonic 
decays of kaons and pions.
Since the angle between the emitted muon and the decaying particle 
(kaon or pion) is small, such a background is suppressed by 
increasing the lower limit on $P_{\chi}$. 

An upper limit on the kaon likelihood ($L_k$) \cite{Staric:1999sk}, 
as measured in the Cherenkov detector, suppresses muon candidates 
from kaon decays.

A lower limit on \pT{} suppresses muons from low mass particle 
decays, since they are expected to have a smaller \pT{} than those 
from $b$ decays.

A lower limit on $Ip$ discriminates muons originating in $b$ decays 
from background muons.
Since the $Ip$ is correlated with the lifetime of the decaying 
particle, and the lifetime of \bfl{} hadrons is larger than that of 
\cfl{} hadrons, the impact parameter cut suppresses open charm 
background.

In order to suppress background in the proximity of the target, a 
lower limit on the difference between the $pmd$ and the target 
positions along the $z$-axis (\Dz) is applied.

Unphysical events are suppressed by the requirement of the following 
upper limits: 
\mbox{$Ip<50\sigma$}, 
\mbox{$\pT<5~\GeVpc$}, 
\mbox{$|\Dz|<5$ cm} and 
\mbox{$m<8~\GeVpcc$}.
The optimisation of the last three cuts (lower limits on \pT, $Ip$ 
and \Dz) is performed simultaneously.

The optimisation procedure for $S$ results in the \bbbar{} selection 
criteria listed in Table \ref{tab:cutopt}, where the number of 
surviving \mpmm{} and \mpmp{} pairs in the data, at each selection 
step, are also shown.
After applying all cuts, the number of remaining dimuon candidates is 

\begin{displaymath}
  \Nmmpm = 167 \pm 13_{stat}.
\end{displaymath}

\vspace{0.3cm}

\begin{table}[htb]
  \center
  \begin{tabular}{|c|c|c|}
    \hline
    Cut                            &    \Nmmpm & \Nmmpp   \\
    \hline
    High quality muon pairs        & $1051593$ & $739947$ \\
    $\pT>1~\GeVpc$, $Ip>1.5\sigma$ & $  34745$ & $ 22359$ \\
    $m\in[2, 8]~\GeVpcc$, no \Jpsi & $  23560$ & $ 19254$ \\
    Upper limits                   & $  23406$ & $ 19129$ \\
    \hline
    $P_{\chi}>0.04$, $L_k<0.9$     & $  16750$ & $ 13268$ \\
    $Ip>4\sigma$                   & $    582$ & $   402$ \\
    $\Dz>0.4$ cm                   & $    167$ & $   100$ \\
    \hline
  \end{tabular}
  \caption{Double muonic $b$ decay selection and surviving 
    \mpmm{} and \mpmp{} pairs in the data obtained from cut optimisation.}
  \label{tab:cutopt}
\end{table}

In the main measurement method, the \Dz{} cut is relaxed to $0.2$ cm 
(see Section \ref{sec:method1}). 
For this cut value, the resulting numbers of opposite-sign and same-sign 
muons are $225$ and $117$, respectively.

\begin{figure}[htb]
  \centerline{\includegraphics[width=0.5\textwidth]
    {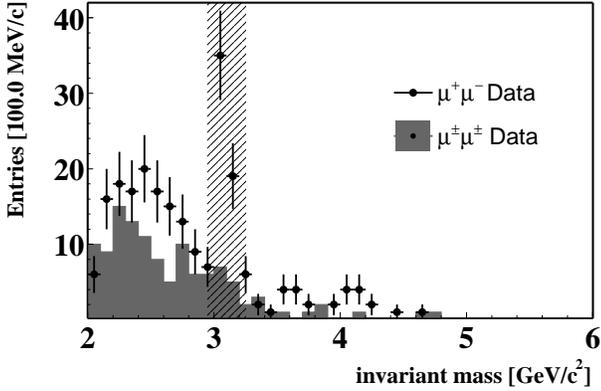}}
  \caption{Dimuon invariant mass after \bbbar{} selection.
    Events in the \Jpsi{} mass region ([$2.95$, $3.25$] \GeVpcc), 
    which is highlighted in the picture, are mostly due to 
    \btojpX{} decays.
    Since they are used in a different analysis \cite{Abt:2005qs}, 
    they are removed.}
  \label{fig:plotmass}
\end{figure}

The invariant mass distribution of dimuon events surviving the 
\bbbar{} selection is shown in Figure \ref{fig:plotmass} (the 
invariant mass cut on the \Jpsi{} is removed for illustrative 
purposes).

\subsection{Cross section determination \label{sec:method1}}

In the main measurement method, the number of \bbbar{} decay events 
in the data (\Nbbpm) is obtained from a likelihood fit to the data of 
the simulated \pT{} and $Ip$ distributions of signal and background 
events.
The selection criteria listed in Table \ref{tab:cutopt} are applied, 
with the exception of the lower limit on \Dz, which is decreased to 
\mbox{$0.2$ cm}.
The number of surviving \mpmm{} pairs becomes 

\begin{displaymath}
  \Nmmpm = 225 \pm 15_{stat}.
\end{displaymath}

The selection on \Dz{} is relaxed because the likelihood fit uses 
more information than that used in the cut optimisation.
The likelihood fit is also sensitive to the shapes of signal and 
background distributions, while the cut optimisation only uses the 
number of signal and background events surviving the selection.

The likelihood function is defined as 

\begin{displaymath}
  L(\Ns, \Nb) = \frac{(\Ns + \Nb)^n e^{{-(\Ns + \Nb)}}}{n!}
  \prod_{i=1}^n \left(\frac{\Ns \Ps + \Nb \Pb}{\Ns + \Nb}\right).
\end{displaymath}
The product index $i$ runs over the $n$ (\mbox{$=\Nmmpm$}) selected 
dimuon events, the exponential term accounts for Poisson fluctuations 
of signal and background, \Ns{} ($=\Nbbpm$) and 
\Nb{} (\mbox{$=\Nccpm+\Ncopm$}) are fit parameters representing the 
number of signal and background events in \mpmm{} data, \Ps{} and 
\Pb{} are the products of the $Ip$ and \pT{} probability 
distributions of the two muons for signal and background events, as 
obtained from the Monte Carlo simulations \cite{Sbrizzi:2006jx}:

\begin{displaymath}
  \begin{array}{l}
    \Ps = \Cs \cdot (\Rs \Pbb + \Pbx)~~~\mathrm{and} \\
    \Pb = \Cb \cdot (\Rb \Pcc + \Pco). \\
  \end{array}
\end{displaymath}

The signal probability (\Ps) is the sum of probabilities from decays 
of class $B1$ (\Pbb), $B2$, $B3$ and $B4$ (\Pbx) with ratio 
\mbox{$\Rs=1$}.
The background probability (\Pb) is obtained by adding the 
probability for charm and combinatorial events (\Pcc{} and \Pco) 
with \mbox{$\Rb=0.2$}.
The probability ratios \Rs{} and \Rb{} are determined through Monte 
Carlo simulations.
Constant factors (\Cs{} and \Cb) are used to normalise the total 
probability to unity.

\begin{figure}[htb]
  \centerline{\includegraphics[width=0.5\textwidth]
    {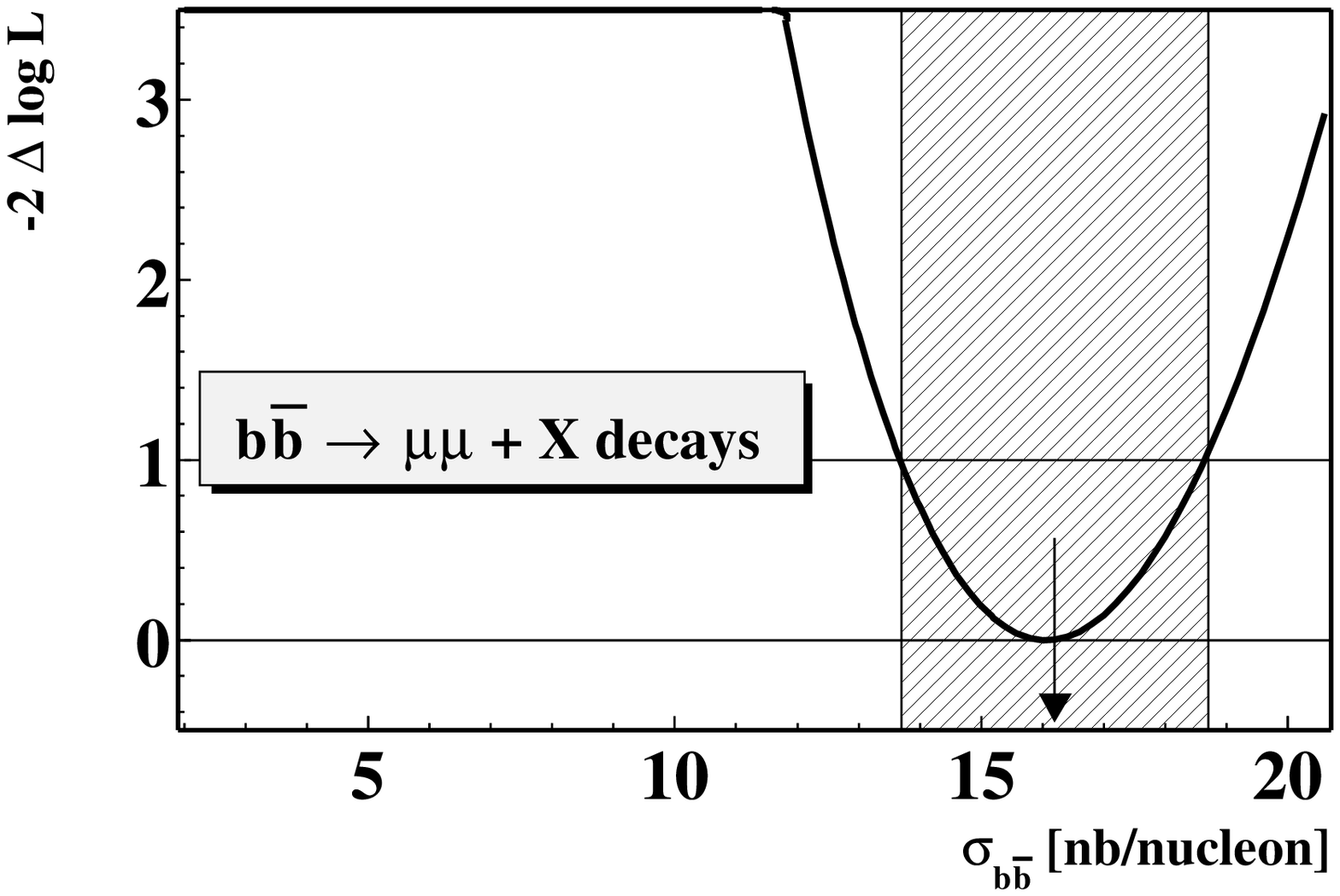}}
  \caption{Dependence of \mbox{$-2\Delta\log{L}=-2(\log{L}-\log{L}_{min})$} 
    on \sigbb, where $L$ is the likelihood of \bbtommX{} decays.
    An increase of one unit on the vertical axis corresponds to a 
    $1\sigma$ variation of \sigbb.}
  \label{fig:lik}
\end{figure}

The result of the likelihood fit performed by minimising the quantity 
$-2\log{L}$ is shown in Figure \ref{fig:lik}.
The minimum is obtained when \mbox{$\Ns=76\pm12$} and 
\mbox{$\Nb=149\pm15$}, which corresponds to the \bbbar{} production 
Cross section:

\begin{displaymath}
  \sigbb = 16.2 \pm 2.5_{stat}~\nbpn.
\end{displaymath}

\subsection{Cross-check by a counting method \label{sec:method2}}

In a measurement method based on event counting, \Nbbpm{} is obtained 
by subtracting all possible background events from the number of 
\mpmm{} pairs surviving the \bbbar{} selection listed in Table 
\ref{tab:cutopt} (\Nmmpm):

\begin{displaymath}
  \Nbbpm = \Nmmpm - \Nccpm - (\Nmmpp - \Nbbpp).
\end{displaymath}

Assuming that the reconstruction efficiency of \bbbar{} decays into 
\mpmm{} is equal to that into \mpmp, for each decay class in each 
target configuration ($\ebbji = \ebbppji = \ebbpmji$), which is true 
to within a few percent, the \bbbar{} production cross section can be 
written as 

\begin{displaymath}     
  \sigbb = \frac{\left(\Nmmpm - \Nmmpp\right) \Dsigjp \BR{\Jpsi} - \sigcc \BR{\ccbar} \Xcc}{\Xbb},
\end{displaymath}
where the term \Xbb{} is given by  

\begin{displaymath}     
  \Xbb = \sum_{i} \Njpi A^{1-\alpha}_{i} \frac{\sum_{j} \BR{Bj} (1-2\thj) \ebbji}{\ejpi}.
\end{displaymath}

The number of dimuon events surviving the \bbbar{} selection are 
\mbox{$\Nmmpm = 167 \pm 13$} and \mbox{$\Nmmpp = 100 \pm 10$}.
The simulated charm background is \mbox{$\Nccpm = 11.7 \pm 0.9$}, 
while the number of simulated \bbbar{} decays into same-sign muons 
are \mbox{$\Nbbpp = 23 \pm 7$}.
The resulting number of \bbbar{} decays into opposite-sign muons is 
\mbox{$\Nbbpm = 78 \pm 23$} (the uncertainty takes into account the 
correlation between \Nbbpm{} and \Nbbpp).

The result of the event counting method is the \bbbar{} cross section 
\mbox{$\sigbb = 18.2 \pm 5.4_{stat}~\nbpn$}. 

Compared to the result of the likelihood fit, the statistical 
uncertainty is increased.
This is due to the fact that the cross section is proportional to the 
difference between two numbers having a large statistical uncertainty 
($\Nbbpm \propto \Nmmpm - \Nmmpp$), while, in the previous method, 
\Nbbpm{} is obtained from \mpmm{} data only.
However, for the likelihood fit, the selection criteria are relaxed, 
which implies that the selected \bbbar{} events contain a larger 
fraction of background.
The advantage is that the likelihood fit takes into account also the 
shapes of the signal and background distributions.

The likelihood fit is used to estimate the number of \bbbar{} events 
in the data, while the event counting method serves to estimate the 
systematic uncertainty associated with the measurement method.

\section{Systematic uncertainty}

Many systematic uncertainties on the measurement presented in this 
paper are similar to those reported in reference \cite{Abt:2005qs}, 
where the reader can find a detailed discussion of detector and 
trigger simulations, models used to simulate \bfl{} hadron production 
and decay, production of \Jpsi{} mesons, fluctuations of the 
proton-nucleus interaction rate, beam position and shape.

The reconstruction efficiency and the production cross section of 
\Jpsi{} mesons are determined assuming no polarisation.
The systematic uncertainty associated with \Jpsi{} production 
includes the effect of a polarization consistent with the limits 
provided by other experiments 
\cite{Alexopoulos:1997yd, Gribushin:1999ha, Chang:2003rz}.

\begin{figure}[htb]
  \centerline{\includegraphics[width=0.5\textwidth]
    {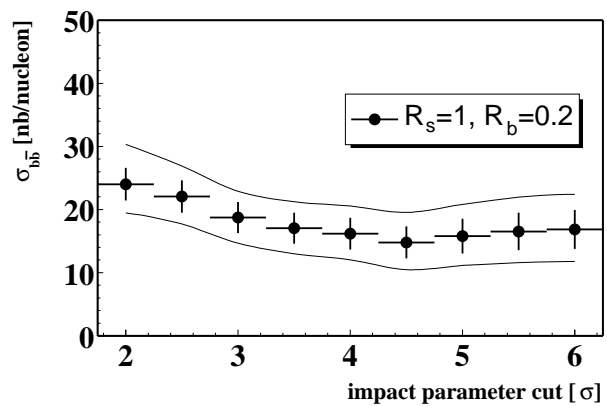}}
  \caption{Dependence of the \bbbar{} production cross section on 
    the muon impact parameter cut.
    The band represents the variation of cross section due to a 
    change of the expected signal and background composition 
    (\Rs{} and \Rb) in the ranges [$0.5, 2$] and [$0, 0.4$], 
    Respectively.}
  \label{fig:sys}
\end{figure}

The contributions specifically affecting the \bbbar{} cross section 
measurement from double muonic $b$ decays are evaluated.
Those which are due to event selection are defined as the maximum
variation of \sigbb, divided by $\sqrt{12}$, obtained when varying 
crucial quantities in the likelihood fit, such as the impact 
parameter cut ($Ip$) and the assumptions on signal and background 
composition (\Rs{} and \Rb).
The $Ip$ cut is varied between $2$ and $6\sigma$, \Rs{} in the range 
[$0.5, 2$] and \Rb{} in the range [$0, 0.4$] (Figure \ref{fig:sys}).

For cut optimisation (Section \ref{sec:cutopt}), a \bbbar{} cross 
section of \mbox{$15~\nbpn$} is assumed.
Since a variation of the value chosen for \sigbb{} might result in a 
different event selection, the systematic uncertainty due to the 
assumption on \sigbb{} is included in the effect of cut variation. 

The systematic effect due to the measurement method is $6\%$.
This effect is estimated by determining \sigbb{} with the two methods 
described in Section \ref{sec:analysis} on the events surviving the 
\bbbar{} selection (Table \ref{tab:cutopt}).

The systematic effect due to the assignment of a priority to the 
classes of double muonic $b$ decays is negligible.

Assuming that all uncertainties listed in Table \ref{tab:syst} are 
independent, the total systematic uncertainty is $18\%$.

\begin{table}[htb]
  \center
  \begin{tabular}{lcc}
   \textbf{Systematic effect}              & \textbf{Uncertainty} \\
    \hline
    Detector and trigger simulation        &    $5\%$ \\
    \Jpsi{} production models              &  $2.5\%$ \\
    \bbbar{} production and decay models   &    $5\%$ \\
    $b$ lifetime                           &    $1\%$ \\
    Proton-nucleus interaction rate        &    $1\%$ \\
    Beam characteristics                   &  $0.5\%$ \\
    \Dsigjp                                &  $8.9\%$ \\
    \Br{\jptomm}                           &  $1.7\%$ \\
    \Jpsi{} nuclear suppression ($\alpha$) &  $3.7\%$ \\
    \Jpsi{} event counting                 &  $0.3\%$ \\
    Efficiency determination               &  $2.0\%$ \\
    Charge factor ($\theta$)               &  $0.3\%$ \\
    \Br{\btomX}                            &  $3.5\%$ \\
    \Br{\ctomX}                            &  $3.2\%$ \\
    \Br{\btoccsX}                          &  $5.5\%$ \\
    $Ip$                                   &    $5\%$ \\
    \Rs{} and \Rb{} ratios                 &    $6\%$ \\
    Measurement method                     &    $6\%$ \\
    \hline
    \textbf{Total}                         &   $18\%$ \\
    \hline
  \end{tabular}
  \caption{List of systematic effects in the measurement of \sigbb.
    The first six contributions are evaluated in reference 
    \cite{Abt:2005qs}.}
  \label{tab:syst}
\end{table}

\section{Conclusions}

The \bbbar{} production cross section in $920$ GeV proton-nucleus 
fixed target collisions has been measured by using double muonic 
\bfl{} hadron decays.
The measurement is performed with a likelihood fit of the simulated 
kinematical distributions of the signal and background events to the 
\HeraB{} dimuon data.
The result is 

\begin{displaymath}
  \sigbb = 16.2 \pm 2.5_{stat} \pm 2.8_{sys}~\nbpn.
\end{displaymath}

The result is consistent with our previous measurement  
\mbox{$\sigbb = 14.9 \pm 2.2_{stat} \pm 2.4_{sys}~\nbpn$} 
\cite{Abt:2005qs}, which was performed on a statistically independent 
set of events.
Both measurements are normalized to the prompt \Jpsi{} production 
cross section.

\begin{figure}[htb]
  \centerline{\includegraphics[width=0.5\textwidth]
    {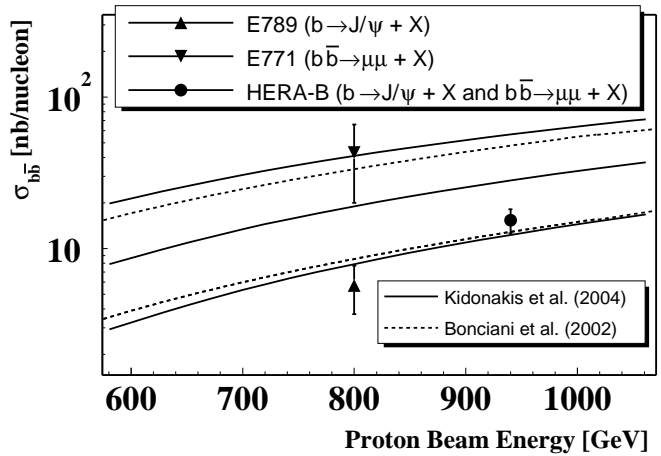}}
  \caption{Cross section for \bbbar{} production as a function of the 
    proton energy in fixed target collisions.
    The predictions of Bonciani~\etal~\cite{Bonciani:1998vc} and 
    Kidonakis~\etal~\cite{Kidonakis:2004qe} are shown.
    The theoretical uncertainties are obtained by changing the 
    renormalisation and factorisation scales and the $b$ mass.
    The \HeraB{} result, which is based on a combined analysis of 
    \btojpX{} and \bbtommX{} decays, is consistent with the 
    theoretical predictions.
    The results of the lower energy Fermilab experiments 
    (E771 \cite{Alexopoulos:1997zx} and E789 \cite{Jansen:1994bz}) 
    are also shown.}
  \label{fig:results}
\end{figure}

The combined result of the two \HeraB{} measurements, accounting for 
the correlation between their systematic uncertainties, is 

\begin{displaymath}
  \sigbb = 15.4 \pm 1.7_{stat} \pm 1.2_{sys}^{uncorr.} \pm 1.9_{sys}^{corr.}~\nbpn,
\end{displaymath}
which is consistent with the latest QCD predictions of 
Bonciani~\etal~\cite{Bonciani:1998vc} and 
Kidonakis~\etal~\cite{Kidonakis:2004qe} 
based on NLO calculations and resummation of soft gluons 
(Figure \ref{fig:results}).

\section*{Acknowledgments}

We thank F. Ferrolini for many stimulating discussions.
We express our gratitude to the DESY laboratory and accelerator group 
for their strong support since the conception of the \HeraB{} 
experiment.
The \HeraB{} experiment would not have been possible without the 
enormous effort and commitment of our technical and administrative 
staff.

  \small
  \bibliographystyle{biblio}
  \bibliography{biblio}

\end{document}